\documentstyle[11pt,epsfig]{article}

\newcommand{\beq}{\begin{equation}}
\newcommand{\eeq}{\end{equation}}
\newcommand{\beqa}{\begin{eqnarray}}
\newcommand{\eeqa}{\end{eqnarray}}

\newcommand{\gae}{\stackrel{>}{\sim}}
\newcommand{\shat}{{\hat s}}
\newcommand{\that}{{\hat t}}
\newcommand{\uhat}{{\hat u}}
\newcommand{\mtsq}{{m_t^2}}

\begin{document}
\begin{titlepage}
\def\thepage {}        

\title{New Strong Interactions at the Tevatron?}

\author{
R.S. Chivukula, A.G. Cohen, and
E.H. Simmons\thanks{e-mail addresses: sekhar@bu.edu, cohen@bu.edu,
simmons@bu.edu},\\
Department of Physics, Boston University, \\
590 Commonwealth Ave., Boston MA  02215}

\date{March 25, 1996}

\maketitle

\bigskip
\begin{picture}(0,0)(0,0)
\put(295,250){BUHEP-96-5}
\put(295,235){hep-ph/9603311}
\end{picture}
\vspace{24pt}

\begin{abstract}
Recent results from CDF indicate that the inclusive cross section for
jets with $E_T > 200$ GeV is significantly higher than that predicted
by QCD.  We describe here a simple flavor-universal variant of the
``coloron" model of Hill and Parke that can accommodate such a jet
excess, and which is not in contradiction with other experimental
data.  As such, the model serves as a useful baseline with which to
compare both the data and other models proposed to describe
the jet excess.  An interesting theoretical feature of the model is that
if the global chiral symmetries of the quarks remain unbroken in the
confining phase of the coloron interaction, it realizes the possibility
that the ordinary quarks are composite particles.

\pagestyle{empty}
\end{abstract}
\end{titlepage}


Recent results from CDF \cite{cdff} indicate that the inclusive cross
section for jets with $E_T > 200$ GeV is significantly higher than
that predicted by QCD. This excess can be fit by a phenomenological
model of quark substructure \cite{cdff}, or by a model containing a
new strongly-coupled $Z'$ gauge boson \cite{hadrophil,altarelli}.
Here we describe a simple flavor-universal variant of the ``coloron"
model of Hill and Parke \cite{coloron} that can accommodate such an
excess, and which is not in contradiction with other experimental
data. The model is minimal in its structure, in that it involves the
addition of one new interaction, one new scalar multiplet, and no new
fermions. As such, the model serves as a useful baseline with which to
compare both the data and other models proposed to describe the
jet excess.  In addition, if the global chiral symmetries of the
quarks remain unbroken in the confining phase of the coloron interaction,
it provides a simple realization of the possibility that the ordinary
quarks are composite particles.

\section{The Model: Higgs Phase Description}
\label{sec:model}
\setcounter{equation}{0}

Following Hill and Parke \cite{coloron}  the QCD
gauge group is extended to $SU(3)_1 \times SU(3)_2$, with gauge couplings
$\xi_1$ and $\xi_2$ respectively, with $\xi_2 \gg \xi_1$.  In contrast
to  Hill and Parke we assign {\it all} quarks to triplet
representations of the strong $SU(3)_2$ group.  As in \cite{coloron},
we break the symmetry to its diagonal subgroup at
a scale $f$ by introducing a scalar-boson $\Phi$ which transforms as a
$(3,\bar{3})$ under the two $SU(3)$ groups. For such a field there are
three independent non-derivative operators of dimension less than or
equal to four, and we can write the potential as
\beq
U(\Phi) = \lambda_1 \mbox{Tr}\left(\Phi\Phi^\dagger - f^2 I\right)^2
          +\lambda_2 \mbox{Tr}\left(\Phi\Phi^\dagger - \frac{I}{3}
            (\mbox{Tr}\,\Phi\Phi^\dagger)\right)^2
\eeq
where we have adjusted the overall constant such that the minimum of
$U$ is zero.  For the range of parameters $\lambda_1,\lambda_2, f^2>
0$ the scalar field $\Phi$ will develop a vacuum expectation value
(VEV), $\langle \Phi \rangle = {\rm diag}(f,f,f)$, breaking the two
strong groups down to a single $SU(3)$ which we identify with QCD.

Once this VEV has developed there remain massless gluons interacting with
quarks  through a conventional QCD coupling  with strength $g_3$, as
well as  an octet of massive colorons ($C^{\mu a}$)  interacting with quarks
through a new QCD-like coupling
\beq
{\cal L} = - g_3 \cot\theta J_\mu^a C^{\mu a}\ ,
\eeq
where $J_\mu^a$ is the color current
\beq
\sum_f \bar{q}_f \gamma_\mu {\lambda^a\over 2} q_f \ ,
\eeq
and where
\beq
\tan\theta = {\xi_1 \over \xi_2}\ .
\eeq
Since $g_3$ is identified with the QCD coupling constant, it has a
value of approximately 1.2 (corresponding to $\alpha_3(M_Z)\approx
.12$). In terms of these parameters the mass of the colorons is
\beq
M_C = \left({g_3 \over \sin\theta \cos\theta}\right) f\ .
\eeq

Below the scale $M_C$, coloron-exchange may be approximated
by the effective four-fermion interaction
\beq
\label{eqn:higgs}
{\cal L}_{eff} = -
{g_3^2\cot^2\theta\over {2! M_C^2}} J_\mu^a J^{\mu a}\ .
\eeq
The effects of this and similar operators on jet production has been
studied\footnote{Following \cite{elp}, in the study of compositeness
it is conventional to define the coefficient of a product of currents
as $4\pi/\Lambda^2$. Using eqn. (\ref{eqn:higgs}), the relationship
between $M_C$ and this conventionally defined $\Lambda$ is $M_C =
\sqrt{\alpha_3}\cot\theta\ \Lambda$ . } in \cite{chosimm}.  Fig. 1
plots the published CDF data \cite{cdff}, the pure leading-order
QCD prediction (corresponding to the limit $M_C \rightarrow \infty$), and the
prediction for\footnote{The analysis of
{\em earlier} (1988-89) CDF jet data \cite{cdfold}, which did not have
a pronounced jet excess at
high-$E_T$, in \cite{chosimm} implies that $M_C/{\cot\theta}$ cannot be much
less
than 700 GeV.} $M_C/\cot\theta = 700$ GeV. As in the case of a contact
interaction between left-handed quarks studied in
\cite{cdff}, the prediction in the presence of this new coloron
interaction is a better fit to the data than the QCD prediction. While
this is suggestive, a complete analysis of this phenomenology and the
assignment of statistical significance requires analysis of the full
data sample.

\begin{figure}
%
\centerline{\epsfig{file=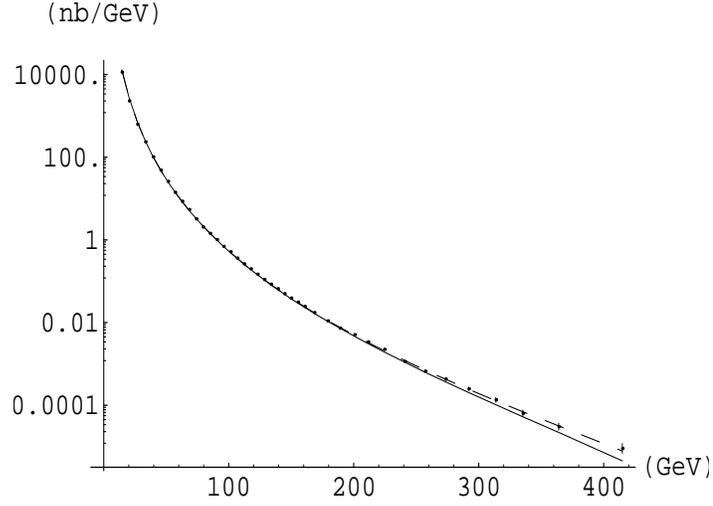, height=7cm,width=10cm}}
%
\vskip .5in
\centerline{\epsfig{file=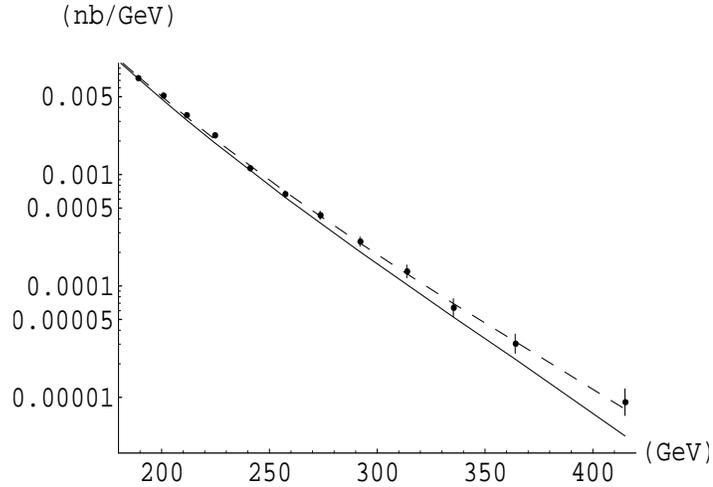,height=7cm,width=10cm}}
\caption[one]{ Single jet inclusive cross-section
 ${1\over{\Delta\eta}} \int (d^2\sigma/d\eta\, dE_T) d\eta $
 as a function of transverse jet energy
 $E_t$, where the pseudorapidity $\eta$ of the jet falls in the range
 $0.1 \leq \vert\eta\vert \leq 0.7$. Dots with (statistical) error bars are
 the recently published CDF data \protect\cite{cdff}.  The solid curve shows
 the LO prediction of pure QCD.  The dashed curve shows the LO
 prediction of QCD plus the color-octet contact interactions of
 equation (\protect\ref{eqn:higgs}) with $M_C/\cot\theta = 700$
 GeV.  Following CDF, we employed the MRSD0' structure functions
 \protect\cite{mrs_pak} and we normalized the curves to the data in
 the region where the effect of the contact interactions
 is small (here this region is $45 < E_T < 95$ GeV).  The upper plot
shows the full transverse-energy range; the lower plot shows more
detail of the high-energy range $E_T > 200$ GeV. }
\label{Single}
\end{figure}

The original coloron model was proposed in the context of ``top-color"
\cite{topcolor} models of electroweak symmetry breaking. The
introduction of the new strong $SU(3)$ gauge-group was motivated by an
attempt to dynamically explain the heavy top quark mass, and
consequently in topcolor models the coloron coupled more strongly to
third generation quarks. The only motivation for the introduction of
the new $SU(3)$ interaction in our model is the potential discrepancy
in the jet data, and the couplings of the coloron in this model are
flavor-universal. As a result, the model can simply be grafted on to
the standard one-doublet Higgs model yielding a simple, complete, and
renormalizable theory\footnote{The most general renormalizable
potential for $\Phi$ and the Higgs-doublet $\varphi$ also includes the
term $\lambda_3 \varphi^\dagger \varphi {\rm Tr}(\Phi^\dagger
\Phi)$. For a range of $\lambda$'s and parameters in the Higgs
potential the vacuum will break the two $SU(3)$ groups to QCD and
break the electroweak symmetry as required.}.

Since the couplings considered here are flavor universal, the theory
is not subject to the customary stringent constraints from flavor
physics \cite{burdman, kominis} in topcolor models.  Coloron
interactions do contribute to corrections to the weak-interaction
$\rho$ parameter (where the isospin splitting is provided through the
$t$ -- $b$ mass splitting). Limits on such corrections
\cite{isospin} imply that
\beq
\frac{M_C}{\cot\theta} \gae 450\: {\rm GeV}.
\eeq
%

\section{Complementarity: The Confining Picture}
\label{sec:complement}
\setcounter{equation}{0}

The model proposed contains scalars in the antifundamental
representation of the strong $SU(3)_2$ gauge group. In the absence of
fermions, such a model exhibits ``complementarity''
\cite{complementarity} with an  exact equivalence between
the Higgs- and the confining-phases. In the presence of quarks
however, there are two possibilities for the physics of the $SU(3)_2$
confining-phase.  The global $SU(6)_L \times SU(6)_R$ chiral
symmetries of the quarks may spontaneously break to $SU(6)_V$ (this is
implicitly assumed to happen in topcolor models).  Alternatively, the
global chiral symmetries may remain unbroken with the ordinary quarks
being massless $SU(3)_2$-singlet composites of the fundamental
fermions and the strongly interacting scalars (much like the
strongly-coupled standard model \cite{scsm}).

If the global chiral symmetries of the quarks remain unbroken in the
$SU(3)_2$ confining phase, this model realizes the possibility that
the ordinary quarks are composite particles.  In this case {\em all}
four-fermion contact interactions consistent with parity and the
global chiral symmetries are allowed. If, in addition to the color
current $J_\mu^a$ defined above, we also define
\beq
J_{5\mu}^a = \sum_f \bar{q}_f \gamma_\mu \gamma_5 {\lambda^a \over 2}
q_f \ ,
\eeq

\beq
J_\mu = \sum_f \bar{q}_f \gamma_\mu q_f\ ,
\eeq
and
\beq
J_{5\mu} = \sum_f \bar{q}_f \gamma_\mu q_f\ ,
\eeq
the most general such interaction may be written
\beq
\label{eqn:confine}
{\cal L} = {4\pi \over {2!\Lambda^2}}\left( c_1 J_\mu^a J^{\mu a}
+ c_2 J_{5\mu}^a J_5^{\mu a} + c_3 J_\mu J^\mu + c_4
J_{5\mu} J_5^{\mu}\right)\ .
\eeq
These terms can be thought
of as arising from the exchange of color-octet and color-singlet,
vector and axial resonances. If $\Lambda$ is chosen to be of
order the masses of these resonances, the $c_i$ are
expected to be of order one \cite{elp,nda}.

\section{Phenomenology}
\label{sec:phen}

The phenomenology of the model depends on whether it is realized in
the Higgs or confining phase. In the Higgs phase, the leading
contribution to new jet physics is due to the exchange of the heavy
coloron,
resulting in the ``VV'' interaction in eqn. (\ref{eqn:higgs}). Away
from the QCD $\hat{t}$-channel pole, this results in an angular
distribution {\it identical} to that of QCD. On the other hand, in the
confining phase one obtains the low-energy interactions
eqn. (\ref{eqn:confine}), with potentially comparable amounts of both
``VV'' and ``AA'' contributions resulting in an angular distribution
which differs from that of QCD.

The angular behavior is implicit in the two-body parton scattering cross
sections
\beq
{d \sigma\over d\that}(ab \to cd) = {\pi \alpha_s^2 \over
\shat^2} \Sigma(ab \to cd).
\eeq
The leading QCD contributions to $\Sigma(ab \to cd)$ may be found in
\cite{Combridgel, Owens} and we have adapted the $O(1/\alpha_s\Lambda^2)$
contributions due to the quark contact operators from
the results of ref. \cite{chosimm}.
The $\Sigma(ab \to cd)$ conventionally include initial state color averaging
factors and are  written in terms of the partonic invariants $\shat$,
$\that$ and $\uhat$.  For scattering of light quarks, whose masses may be
neglected, we have:
\begin{eqnarray}
\Sigma(qq'\to qq') &=& {4 \over 9} {\shat^2+\uhat^2 \over \that^2}
+{8 \over 9} {(c_1+c_2)\shat^2+(c_1-c_2)\uhat^2 \over \alpha_s
\Lambda^2\, \that} \\
&&+{4 \over 9}\left[{\big(2(c_1+c_2)^2+9(c_3+c_4)^2\big)\shat^2
+\big(2(c_1-c_2)^2+9(c_3-c_4)^2\big)\uhat^2\over2\alpha^2_s\Lambda^4}\right]
\nonumber\\
&&+ O\left({1 \over\alpha_s \Lambda^4}\right)  \nonumber\\
&&\nonumber \\
\Sigma(q\bar{q}\to q' \bar{q}') &=& {4\over 9} {\that^2+\uhat^2 \over \shat^2}
+{8 \over 9} {(c_1+c_2)\uhat^2+(c_1-c_2)\that^2 \over \alpha_s
\Lambda^2\, \shat} \\
&&+{4 \over 9}\left[{\big(2(c_1+c_2)^2+9(c_3+c_4)^2\big)\uhat^2
+\big(2(c_1-c_2)^2+9(c_3-c_4)^2\big)\that^2\over2\alpha^2_s\Lambda^4}\right]
\nonumber\\
&&+ O\left({1 \over\alpha_s \Lambda^4}\right)  \nonumber\\
&&\nonumber \\
\Sigma(qq \to qq) &=& {4 \over 9} \left( {\shat^2+\uhat^2 \over \that^2}
+{\shat^2+\that^2 \over \uhat^2} \right)
-{8\over 27} {\shat^2\over \that \uhat}  \\
&&+{8 c_1 \over 9\alpha_s\Lambda^2} \left({\shat^2+\uhat^2 \over \that}
+{\shat^2+\that^2 \over \uhat} \right)
+{8 c_2 \over 9\alpha_s\Lambda^2} \left({\shat^2-\uhat^2 \over \that}
+{\shat^2-\that^2 \over \uhat} \right)  \nonumber \\
&&+\left({8 (c_1+c_2)\over 27 \alpha_s\Lambda^2}-{16 (c_3+c_4)\over
9\alpha_s\Lambda^2} \right) {\shat^3\over \that \uhat} \nonumber \\
&&+{4 \over 9}\left[{\big(2(c_1+c_2)^2+9(c_3+c_4)^2\big)2\shat^2
+\big(2(c_1-c_2)^2+9(c_3-c_4)^2\big)(\uhat^2+\that^2)
\over2\alpha^2_s\Lambda^4}\right] \nonumber \\
&& -{8\shat^2\over 27 \alpha^2_s\Lambda^4}
\left[(c_1 + c_2 -6c_3 -6c_4)^2-{81\over 2}(c_3+c_4)^2\right] \nonumber \\
&&+ O\left({1 \over\alpha_s \Lambda^4}\right)  \nonumber\\
\Sigma(q\bar{q}\to q\bar{q}) &=& {4\over 9}\left({\shat^2+\uhat^2\over\that^2}
+{\that^2+\uhat^2 \over \shat^2} \right)
-{8\over 27} {\uhat^2\over \shat\that}  \\
&& +{8 c_1 \over 9\alpha_s\Lambda^2} \left({\shat^2+\uhat^2 \over \that}
+{\that^2+\uhat^2 \over \shat} \right)
-{8 c_2 \over 9\alpha_s\Lambda^2} \left({\shat^2-\uhat^2 \over \that}
+{\that^2-\uhat^2 \over \shat} \right) \nonumber \\
&& +\left({8  (c_1+c_2)\over 27\alpha_s \Lambda^2}-{16 (c_3+c_4)\over
9\alpha_s\Lambda^2}
\right) {\uhat^3\over \shat\that} \nonumber \\
&&+{4 \over 9}\left[{\big(2(c_1+c_2)^2+9(c_3+c_4)^2\big)2\uhat^2
+\big(2(c_1-c_2)^2+9(c_3-c_4)^2\big)(\shat^2+\that^2)
\over2\alpha^2_s\Lambda^4}\right] \nonumber \\
&& -{8\uhat^2\over 27 \alpha^2_s\Lambda^4}
\left[(c_1 + c_2 -6c_3 -6c_4)^2-{81\over 2}(c_3+c_4)^2\right] \nonumber \\
&&+ O\left({1 \over\alpha_s \Lambda^4}\right)
\nonumber
\end{eqnarray}
where $q'$ denotes a quark of a flavor other than $q$.   The top quark
is heavy enough that it must be treated separately.  Firstly the top
quark content of the proton is negligible and we need not consider
contributions from initial-state top quarks.  Secondly top quarks are
produced in quark-quark scattering with a mass dependent cross-section
\cite{Combridgell, chosimm}
\begin{eqnarray}
\Sigma(q \bar{q} \to t \bar{t})  &=& {4 \over 9 \shat^2}
  \left[ \that^2+\uhat^2+4\mtsq\shat-2 m_t^4 \right] \\
&& + {8\over 9 \shat \alpha_s \Lambda^2}\left[
c_1(\that^2+\uhat^2+4\mtsq\shat-2 m_t^4)
  + c_2 \shat (\that-\uhat) \right] \nonumber \\
&& + { 4 \over 9 \alpha^2_s \Lambda^4} \Big[({c_1}^2+ \frac 92 {c_3}^2)
  (\that^2+\uhat^2+4\mtsq\shat-2 m_t^4) \nonumber \\
&& \quad\qquad + ({c_2}^2 +\frac 92 {c_4}^2)
  (\that^2+\uhat^2-2 m_t^4 ) \nonumber \\
&&  \quad\qquad + (2c_1 c_2 + 9 c_3 c_4) \shat(\that-\uhat)
\Big].
\end{eqnarray}

Since we are including terms that are $O(1/\Lambda^4)$ in the
scattering cross-sections, we need to comment on possible
contributions from dimension-8 operators.  The dimension-eight
operators that contribute to the processes above include two more
derivatives than the dimension-6 operators in (\ref{eqn:confine});
for instance, one of the operators is
\beq
{4\pi\over{2!\Lambda^4}} D^\nu J^\mu_a D_\nu J_{\mu\,a}.
\eeq
The contributions this operator makes to the scattering amplitude will
clearly of the same form as those of the related dimension-6 operator
-- but will be suppressed by a factor of $s/\Lambda^2$, as
one would expect from the rules of dimensional analysis \cite{nda}.
The leading contributions of such dimension-8 operators to the
cross-section (which arise from interference with QCD) are
$O(1/\alpha_s\Lambda^4)$, {\it i.e.} down by $O(\alpha_s)$
relative to the contributions from the dimension-6 operators kept
above.

It is important to note that the high-$E_T$ jet excess predicted by this model
will be {\it flavor universal}. Regardless of whether the model is in the
Higgs or confining phase, the characteristics (rate and angular
distribution) of jets at high-$E_T$ should be the same for jets with
tagged $b$- or $c$-quarks as for all quark-jets.

Finally, at higher energy hadron colliders such as the LHC one would
see (potentially broad) resonances in the two-jet cross section at
invariant masses of order one to several TeV. In the Higgs phase,
the resonances would correspond to colorons, while in the confining
phase one would expect color-octet and color-singlet,
vector and axial bound state resonances.

\section{Conclusions and Caveats}
\label{sec:caveat}

In this note we have described a simple flavor-universal variant of
the coloron model of Hill and Parke that can accommodate the
apparent excess of high-$E_T$ jets at the Tevatron. The model is
minimal in its structure, in that it involves the addition of one new
interaction, one new scalar multiplet, and no new fermions. As such,
the model serves as a useful baseline with which to compare both the
data and other models proposed to describe the jet excess.
Furthermore, if the global chiral symmetries of the quarks remain
unbroken in the confining phase of this new interaction, it provides a
simple realization of the possibility that the ordinary quarks are
composite particles.

Theoretically, the biggest draw-back of this model is that it
introduces new physics at an energy scale of order a TeV without
contributing to an explanation of electroweak or flavor symmetry
breaking. If features of this model are confirmed, it is to be hoped
that the actual dynamics is based on an extension of the model that
will bear on these questions.  For example, some ``Composite
Technicolor Standard Models'' \cite{ctsm} contain chiral coloron gauge
groups (which are used to break the extended technicolor gauge
symmetries) and produce flavor-universal, though not parity-invariant,
interactions of a form similar to eqn. (\ref{eqn:confine}).

\bigskip
\centerline{\bf Acknowledgments}
\vspace{12pt}

We thank Ken Lane, John Terning, Peter Cho and Bogdan Dobrescu for
useful conversations and critical discouragement. R.S.C. thanks
C. Verzegnassi and the Theory Group of the University of Lecce for
their hospitality during the completion of this work.
E.H.S.  acknowledges the support of an NSF Faculty Early Career
Development (CAREER) award and an DOE Outstanding Junior Investigator
award. {\em This work was supported in part by the National Science
Foundation under grants PHY-9057173 and PHY-9501249, and by the
Department of Energy under grant DE-FG02-91ER40676.}

\newpage



\begin{thebibliography}{99}
\frenchspacing

\bibitem{cdff} ``Inclusive Jet Cross Section in $\bar{p}p$ Collisions
at $\sqrt{s} = 1.8$ TeV'', CDF Collaboration, F.~Abe {\it et. al.},
FERMILAB-PUB-96/020-E, hep-ex/9601008.

\bibitem{hadrophil} ``Hadrophilic $Z'$: A Bridge from LEP1, SLC
and CDF to LEP2 Anomalies'', P.~Chiapetta, J.~Layssac, F.~M.~Renard,
and C.~Verzegnassi, hep-ph/9601306.

\bibitem{altarelli} ``$R_b$, $R_c$, and Jet Distributions at the
Tevatron in a Model with an Extra Vector Boson'', G.~Altarelli,
N.~Di~Bartolomeo, F.~Feruglio, R.~Gatto, and M.~L.~Mangano,
hep-ph/9601324.

\bibitem{coloron} C.~T.~Hill and S.~J.~Parke, Phys. Rev. {\bf D49}
(1994) 4454.

\bibitem{chosimm} E.~H.~Simmons, Phys. Lett. {\bf B226} (1989) 132 and
Phys. Lett. {\bf B246} (1990) 471; P.~Cho and E.~H.~Simmons
Phys. Lett. {\bf B323} (1994) 401 and Phys. Rev. {\bf D51} (1995) 2360.

\bibitem{elp}  E.~Eichten, K.~Lane, and M.~E.~Peskin,
Phys. Rev. Lett. {\bf 50} (1983) 811.

\bibitem{cdfold} CDF Collaboration, F. Abe {\it et al.},
FERMILAB-PUB-91/231-E\ semi
CDF Collaboration, F. Abe {\it et al.}, Phys. Rev. Lett. {\bf 68}
 (1992) 1104.

\bibitem{mrs_pak} A.D. Martin, R.G. Roberts and W.J. Stirling,
Phys. Lett. {\bf B306} (1993) 145.

\bibitem{topcolor} C.~T.~Hill, Phys. Lett. {\bf B266} (1991) 419.

\bibitem{burdman} ``GIM Violation and New Dynamics of the Third
Generation'', G.~Buchalla, G.~Burdman, C.~T.~Hill, and D.~Kominis,
FERMILAB-PUB-95-322-T and hep-ph/9510376.

\bibitem{kominis} D.~Kominis, Phys. Lett. {\bf B358} (1995) 312.

\bibitem{isospin} R.~S.~Chivukula,  B.~A.~Dobrescu, and
J.~Terning,  Phys. Lett. {\bf B353} (1995) 289.


\bibitem{complementarity} G.~'t~Hooft, Nucl. Phys. {\bf B138}
(1978) 1;  E.~Fradkin and S.~H.~Shenker, Phys. Rev. {\bf D19}
(1979) 3682; T.~Banks and E.~Rabinovici,  Nucl. Phys.
{\bf B160} (1979) 349;  S.~Dimopoulos, S.~Raby, and
L.~Susskind, Nucl. Phys. {\bf B173} (1980) 208.

\bibitem{scsm} L.~Abbott and E.~Farhi,  Phys. Lett. {\bf 101B} (1981)
69 and Nucl. Phys. {\bf B189} (1981) 547.

\bibitem{nda} A.~Manohar and H.~Georgi,  Nucl. Phys.
{\bf B234} (1984) 189;
H.~Georgi and L.~Randall, Nucl. Phys. {\bf B276} (1986) 241;
R.~S.~Chivukula, M.~J.~Dugan, M.~Golden, Phys. Rev. {\bf D47}
(1993) 2930.

\bibitem{Combridgel} B.L. Combridge, J. Kripfganz and J. Ranft, Phys. Lett.
{\bf B70} (1977) 234.

\bibitem{Owens} J.F. Owens and E. Reya, Phys. Rev. {\bf D18} (1978) 1501.

\bibitem{Combridgell} B.L. Combridge, Nucl. Phys. {\bf B151} (1979) 429.

\bibitem{ctsm} R.~S.~Chivukula and H.~Georgi, Phys. Lett. {\bf B188}
(1987) 99; R.~S.~Chivukula, H.~Georgi, and L.~Randall, Nucl. Phys.
{\bf B292} (1987) 83; R.~S.~Chivukula and H.~Georgi, Phys.
Rev. {\bf D36} (1987) 102.

\end{thebibliography}
\end{document}